\documentclass[twocolumn,twocolappendix]{aastex701}

\newcommand{\y}[1]{{\textcolor{red}{#1}}}

\usepackage{caption}

\hypersetup{colorlinks,%
 citecolor=blue,%
 linkcolor=blue}
\newcommand{\kms}{{\rm km}\,{\rm s}^{-1}{}}

\shorttitle{Tidal Grinding of Dwarf Ellipticals}
\shortauthors{Paudel et al.}

\begin{document}
\title{Tidal Grinding of Dwarf Galaxies in Cluster Environments
}

\author[orcid=0000-0003-2922-6866,gname='Sanjaya',sname='Paudel']{Sanjaya Paudel}
%\altaffiliation{Department of Astronomy, Yonsei University, Seoul 03722, Republic Of Korea}

\affiliation{Department of Astronomy \& Center for Galaxy Evolution Research, Yonsei University, Seoul 03722, Republic Of Korea }
\email[hide]{sanjpaudel@google.com}  
\affiliation{Nepal Astronomical Society, Kathmandu, Nepal}
%\author[orcid=0000-0002-5513-5303,gname='Cristiano G.', sname='Sabiu']{Cristiano G. Sabiu} 
%\altaffiliation{Las Campanas Observatory}
%\affiliation{Natural Science Research Institute (NSRI), University of Seoul, Seoul 02504, Republic of Korea}
%\email{csabiu@google.com}

\author[orcid=0000-0002-1842-4325,sname='Yoon']{Suk-Jin Yoon}
\affiliation{Department of Astronomy \& Center for Galaxy Evolution Research, Yonsei University, Seoul 03722, Republic Of Korea }
\email[show]{sjyoon0691@yonsei.ac.kr}

%\author[orcid=0000-0002-1842-4325,sname='Yoon']{Et. al}
%\affiliation{Department of Astronomy \& Center for Galaxy Evolution Research, Yonsei University, Seoul 03722, Republic Of Korea }
%\email[hide]{sjyoon0691@yonsei.ac.kr}

\author[orcid=0000-0003-4586-0744,sname='Adhikari']{Tek Prasad Adhikari}
\affiliation{CAS Key Laboratory for Research in Galaxies and Cosmology, Department of Astronomy, University of Science and Technology of China, Hefei, Anhui 230026, China}
\affiliation{School of Astronomy and Space Science, University of Science and Technology of China, Hefei, Anhui 230026, China}
\email{tek@ustc.edu.cn}

\author[orcid=0009-0004-2083-3111,sname='Gim']{Eun-Taek Gim}
\affiliation{Department of Astronomy \& Center for Galaxy Evolution Research, Yonsei University, Seoul 03722, Republic Of Korea }
\email[hide]{kut1996@yonsei.ac.kr}

\author[orcid=0009-0008-4928-0599,sname='Kim']{Myung-Hun Kim}
\affiliation{Department of Astronomy \& Center for Galaxy Evolution Research, Yonsei University, Seoul 03722, Republic Of Korea }
\email[hide]{kut1996@yonsei.ac.kr}

\author[orcid=0009-0007-7069-4414,sname='Park']{Inhyuk Park}
\affiliation{Department of Astronomy \& Center for Galaxy Evolution Research, Yonsei University, Seoul 03722, Republic Of Korea }
\email[hide]{kut1996@yonsei.ac.kr}

\author[orcid=0000-0003-2569-8129,sname='Pokhrel']{Nau Raj Pokhrel}
\affiliation{Department of Physics and Astronomy, The University of Tennessee, Knoxville, TN 37996, USA}
\email{npokhrel@utk.edu}

\begin{abstract}
Dwarf elliptical galaxies (dEs) dominate galaxy clusters and provide key constraints on environmentally driven galaxy evolution. Here we examine whether the projected shapes of dEs retain information about their accretion and transformation histories using a homogeneous sample of 1,108 bright ($m_g < 19$ mag) dEs in the Virgo cluster.
Based on the axis-ratio ($b/a$), we define flat ($< 0.70$) and round ($> 0.74$) subsamples and compare their spatial and kinematic properties. We find that flat dEs are distributed more uniformly across the cluster, whereas round dEs preferentially occupy regions of stronger tidal fields around massive ($M_* > 10^{10}\,{\rm M}_\odot$) galaxies.
Within the central $5^\circ \times 5^\circ$ region around the Virgo central galaxy (M87), 149 dEs have spectroscopic radial velocities compiled from public archives. 
In this region, the two shape classes also exhibit clear kinematic segregation: flat dEs have systematically larger line-of-sight velocity offsets from the cluster mean (median $\Delta v = 654~\kms$), whereas round dEs have smaller offsets (median $\Delta v = 414~\kms$), as expected for a more dynamically relaxed population. Flat dEs are consistent with a population that has experienced weaker tidal processing and consequently retains more flattened morphologies. By contrast, round dEs are consistent with prolonged tidal processing (``tidal grinding'') that may have transformed initially flattened systems into rounder spheroids. However, projection contamination of the round subsample may have introduced some uncertainty in the interpretation of intrinsic galaxy shapes.

\end{abstract}


\section{Introduction}
The shapes of dwarf galaxies encode their dynamical histories and provide a sensitive probe of environmental processing in galaxy clusters \citep{Lisker07,Lisker09,Vijayaraghavan15,Carlsten21,Janz14}. Dwarf galaxies are the most numerous galaxy population in the Universe and are highly sensitive to environmental processes \citep{Binggeli85,Rekola05,Tolstoy09}. In galaxy clusters, the shallow gravitational potentials of dwarfs make them particularly vulnerable to tidal interactions, encounters with other galaxies, and the hot intracluster medium. These processes can significantly alter their structure, stellar populations, and kinematics, potentially transforming rotationally supported systems into pressure-supported spheroidal galaxies \citep{Moore96,Mayer01,Mastropietro05,Boselli06,Kazantzidis11}. Understanding how cluster environments influence the structural evolution of dwarf galaxies, therefore, provides important insight into the role of environment in low-mass galaxy evolution.

Nearby clusters such as the Virgo Cluster and Fornax Cluster exhibit a strong environmental dependence of galaxy morphology, commonly referred to as the morphology–density relation \citep{Dressler80,Postman84,Goto03}. While this relation is well established for massive galaxies, its origin and manifestation among dwarf galaxies remain less clear. In particular, the intrinsic shapes and spatial distributions of dwarfs can provide key constraints on their dynamical histories and the timescales over which environmental processes operate \citep{Ferguson89,Janssen10,Lisker18}.

Early-type dwarf galaxies (dEs), which are predominantly found in clusters and groups, thought to be subject to environmental transformation processes \citep{Ferguson94,Boselli06,Lisker07}. Within clusters, their abundance increases toward regions of higher local galaxy density and they often cluster around massive early-type galaxies \citep{Dressler80,Binggeli87,Tully08}. These trends are commonly interpreted as the result of mechanisms such as ram-pressure stripping and tidal interactions that transform late-type progenitors into quiescent systems \citep{vanZee04,Sabatini05,Boselli08}. However, detailed studies reveal that dEs are structurally diverse: many show disk features and axial-ratio distributions consistent with thick disks rather than purely spheroidal systems \citep{Lisker06, Lisker07}, indicating that multiple evolutionary pathways may contribute to their formation.

In this Letter, we investigate the spatial distribution of dEs as a function of their apparent axis ratio within the nearby Virgo Cluster. Previous studies have suggested that the structural properties of dEs may be linked to their dynamical histories within the cluster environment. In particular, \citet[][hereafter~L09]{Lisker09} reported that dEs with different apparent shapes exhibit distinct kinematic properties relative to the cluster potential: low axis-ratio (flattened) systems tend to have higher velocities with respect to the cluster mean, whereas high axis-ratio (rounder) dEs move more slowly and are likely more dynamically relaxed. However, that analysis was limited to only 49 dwarf galaxies with available radial velocities located primarily in the central region of the Virgo cluster.

Here we extend this investigation to the entire extended Virgo cluster region, covering an area of approximately $12\arcdeg\times16\arcdeg$. By analyzing the spatial distribution of a much larger sample of dEs across the cluster environment, we examine whether the morphological segregation suggested by previous kinematic studies is also reflected in the large-scale spatial distribution of dwarf galaxies.

\section{Sample selection}
The dE sample used in this work was identified in our previous studies \citep{Paudel25,Paudel25b} using a deep-learning detection pipeline that combines a region-based convolutional neural network (R-CNN) with an External Attention Network (EANet) classifier, applied to wide-field Legacy Survey imaging of the Virgo Cluster. The classifier was trained on a visually inspected catalog of more than 5,000 dEs spanning a wide range of environments (clusters, groups, and fields), drawn from our earlier systematic search of nearby-Universe ($z \leq 0.01$) Legacy Survey imaging. Validated against the independent, deep Next Generation Virgo Cluster Survey (NGVS) catalog in the Virgo core, the method recovers $\sim81\%$ of dEs at $m_g < 20$ mag.

As the nearest massive galaxy cluster, Virgo provides an ideal laboratory for studying the environmental processes that shape galaxy morphology. Located at a distance of approximately 16.5 Mpc, the cluster has a virial mass of about and contains a rich population of galaxies spanning a wide range of stellar masses, from giant ellipticals to ultra-faint dwarfs \citep{Binggeli85,Mei07,Ferrarese16}. Its relatively close proximity allows detailed structural measurements of low-mass galaxies that are difficult to obtain in more distant clusters.

\begin{figure}
\includegraphics[width=8.5cm]{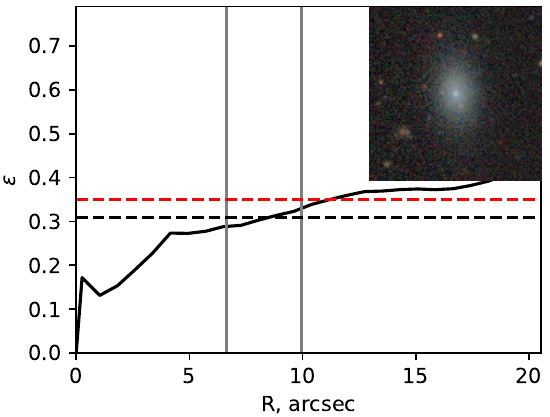}
\caption{Major-axis ellipticity profile of an example dE. The ellipticity profile is derived from isophotal ellipse fitting to the galaxy light distribution. The two vertical gray lines mark 0.75 and 1.25 times the half-light radius of the galaxy, defining the radial range used to measure the representative axis ratio. The horizontal black dashed line indicates the median ellipticity measured within this range. For comparison, the red horizontal dashed line shows the ellipticity reported by the Legacy Survey pipeline. A combined $g–r–z$ tricolor image of the galaxy is shown in the upper-right corner.}
\label{elifit}
\end{figure}

The Virgo Cluster also exhibits a complex dynamical structure, including multiple subclusters and a dense core dominated by massive galaxies such as M87 and M49 \citep{Binggeli87,Schindler99}. This diversity of environments—from the high-density cluster center to lower-density outskirts—makes Virgo particularly well suited for investigating how galaxy morphology depends on environment. In particular, its large population of dwarf galaxies provides a unique opportunity to examine how structural properties, such as galaxy shape and axis ratio, vary with cluster-centric location and proximity to massive galaxies.

\subsection{Measurement of axis ratio}
To ensure reliable structural measurements, we restrict our analysis to relatively bright dEs with apparent magnitudes $m_{g}\,<\,19$ mag (for the average distance of Virgo cluster 16.5 Mpc, this is equivalent to $M_{g} = -12.01$ mag or M$_{*} \approx 10^{7} M_{\sun}$  ). At this brightness level, the signal-to-noise ratio of the galaxy images is sufficiently high to allow robust determination of structural parameters such as ellipticity and axis ratio.

\begin{figure}
\includegraphics[width=8.5cm]{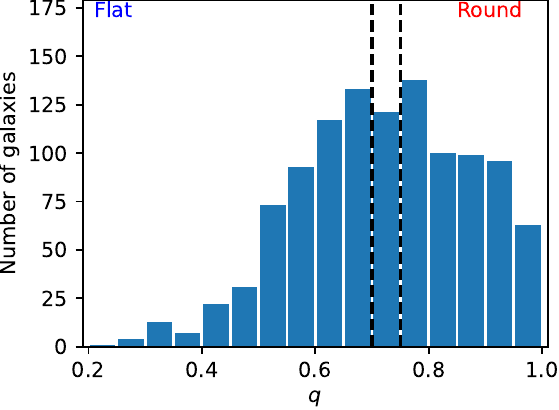}
\caption{Distribution of axis ratios for the full sample of dEs. The median axis ratio of the sample is $q=0.72$. The vertical dashed lines mark the 45th and 55th percentiles of the distribution, which define the boundaries used to classify galaxies by shape. dEs with $q<0.70$ are defined as flat, while those with $q>0.74$ are defined as round. The intermediate axis ratios between these percentiles are excluded to avoid ambiguous classifications and to ensure a clean separation between flattened and round systems.  }
\label{axrhist}
\end{figure}

The axis ratios of the dwarf galaxies were measured using  $g-$band imaging from the Legacy Imaging Surveys\footnote{https://www.legacysurvey.org/} \citep{Dey19}. For each galaxy, we retrieved stacked  $g-$band image cutouts with a size of $2\arcmin\times2\arcmin$  from the survey archive. This image size provides adequate spatial coverage of the galaxy and its immediate surroundings, allowing accurate background estimation and masking of contaminating sources.

Sky background subtraction was performed individually for each galaxy field, as we have in our previous work.  Sky background maps were constructed by masking all detected sources using \texttt{SExtractor} segmentation maps, filling masked regions with surrounding median pixel values, and subtracting the resulting background model from the original image, see \citet{Paudel23,Paudel25} for detail.

We derived the structural parameters of each galaxy using an ellipse-fitting procedure implemented in the photutils package within the Python astropy ecosystem. The algorithm fits a series of concentric elliptical isophotes to the two-dimensional galaxy light distribution following the method of \cite{Jedrzejewski87}. Surface brightness is sampled along ellipses of increasing semi-major axis while the center and position angle are held fixed and the ellipticity, $\epsilon$, is allowed to vary. This procedure yields the major-axis surface-brightness profile together with the radial variation of ellipticity. This approach minimizes instabilities in the fit while allowing the intrinsic shape of the galaxy isophotes to be captured.

The final axis ratio, $b/a= 1-\epsilon$, of each galaxy was determined by averaging the ellipticity measurements over a radial range between 
0.75 and 1.25 times the half-light radius. This radial interval was chosen to avoid the central region, where seeing effects and nuclear components may bias the shape measurement, while also excluding the outermost low signal-to-noise regions of the galaxy. The half-light radii used for this analysis were taken from our previous structural measurements performed during the construction of the main galaxy catalog.

The distribution of measured projected axis ratios, $q = b/a$, is shown in Figure~\ref{axrhist}. To investigate how the spatial and kinematic properties of dwarf ellipticals depend on intrinsic shape, we divide the sample into two morphological subclasses based on the axis-ratio distribution. Galaxies with $q<0.70$ (below the 45th percentile) are classified as \textit{flat} dEs, while those with $q > 0.74$ (above the 55th percentile) are classified as \textit{round} dEs. This percentile-based threshold naturally excludes galaxies with intermediate axis ratios, producing two subsamples of comparable size while minimizing contamination from systems whose intrinsic shapes are ambiguous.

Of the 1,108 dwarf ellipticals in our parent sample, 499 are classified as flat ($q<0.70$) and 499 as round ($q>0.74$), with the remaining galaxies occupying the intermediate axis-ratio range. These thresholds correspond to the 45th and 55th percentiles of the observed axis-ratio distribution and were chosen to exclude systems with intermediate projected shapes while maintaining two subsamples of equal size. We emphasize that these cuts do not represent a physically distinct boundary between galaxy populations but provide a convenient framework for comparing the extremes of the observed shape distribution. We verified that the main trends remain qualitatively unchanged when alternative axis-ratio thresholds are adopted, see Table 1.

\section{results}

%%\begin{figure}
\begin{figure*}
\includegraphics[width=16.7cm]{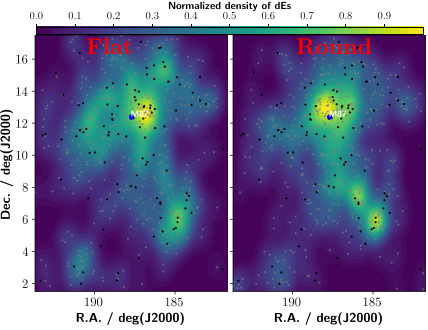}
\caption{Surface number density maps of dEs in the Virgo cluster. Flat dEs are shown in the left panel and round dEs in the right panel. The maps are constructed using a two-dimensional Gaussian kernel density estimation (KDE) with a kernel width of 0\rlap{.}$^{\circ}$175 = 50 kpc. The color bar indicates the normalized surface density. Small gray points represent individual dEs used to generate the density maps, while large black circles mark massive galaxies.}
\label{classmap}
%%\end{figure}
\end{figure*}

Figure~\ref{classmap} presents the projected spatial distribution of flat and round dEs across the Virgo Cluster, revealing clear differences in their cluster-wide distributions. Round dEs are preferentially concentrated around the massive (M$_{*} > 10^{10}$) galaxies of the Virgo subclusters tracing the strong tidal environment. In contrast, flat dEs display a more spatially extended and homogeneous distribution across the cluster, exhibiting only a single dominant concentration near the cluster core surrounding M~87. 
This distinction suggests that round and flat dEs have experienced systematically different degrees of environmental processing, consistent with round dEs residing preferentially near massive cluster members, where stronger tidal forces can reshape their structure over time.

To quantify this difference statistically, we performed a two-dimensional two-sample Kolmogorov–Smirnov (2D KS) test on the projected sky coordinates of the two populations \citep{Peacock83}. We employed the algorithm of \citet{Fasano87}, which generalizes the classical one-dimensional KS test to two dimensions by computing the maximum difference between the cumulative distribution functions across all four natural quadrants defined at each data point. We estimated the significance empirically through 1000 Monte Carlo permutations, in which the spatial labels of the two populations were randomly shuffled to build the null distribution of the test statistic D.

Applying this test to the flat and round dwarf populations yields $D = 0.20$ and $p = 0.002$, indicating that the two spatial distributions are statistically distinct at the $\approx 3.1\sigma$ level. We therefore reject the null hypothesis that flat and round dEs are drawn from the same underlying projected distribution at $99.8\%$ confidence.

\begin{table}
\caption{Results of the two-dimensional KS tests obtained using different percentile-based axis-ratio cuts to define the flat and round dE subsamples.}
\begin{tabular}{c|c|cc}
\hline
Percentile & q & P & D \\
\hline
45/55 & 0.70/0.74 & 0.002 & 0.20 \\
\hline
35/65 & 0.65/0.78 & 0.006 & 0.12\\
\hline
25/75 & 0.61/0.84 & 0.002 & 0.19\\
\hline
\end{tabular}
\tablecomments{The first column lists the adopted percentile thresholds and the second column the corresponding q limits. The third and fourth columns give the KS-test probability and KS statistic, respectively. The consistency of the results across a range of thresholds demonstrates that the observed environmental segregation is not sensitive to the specific axis-ratio cut adopted in this work. }
\end{table}

To assess the robustness of this result against the adopted axis-ratio thresholds, we repeated the analysis using a range of alternative definitions for the flat and round subsamples. We find that the spatial segregation persists for all reasonable choices of the axis-ratio cuts, with comparable KS statistics despite variations in sample size. Table~1 summarizes the results of the 2D KS tests for several alternative selection criteria, demonstrating that our conclusions are not driven by the specific thresholds adopted in this work.

\begin{figure}
\includegraphics[width=8.5cm]{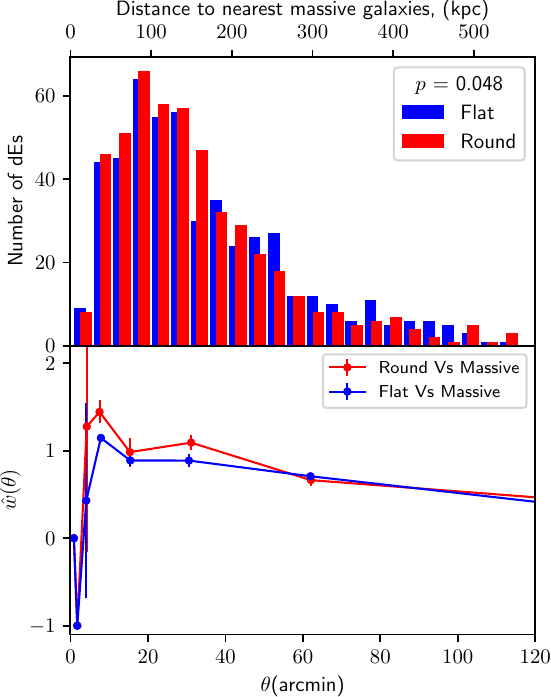}
\caption{Top: Distribution of projected distances from each dEs to its nearest massive galaxy. The $p$-value from the KS test comparing the flat and round dE distributions is indicated in the legend.\\ 
Bottom: Projected two-point cross-correlation functions between dEs and massive galaxies. Round dEs exhibit a stronger clustering signal at small projected separations compared to flat dEs. Error bars are derived from jackknife resampling.}
\label{radprof}
\end{figure}

To quantify the spatial association between dwarf galaxies and massive cluster members, we compute the projected cross-correlation function between dEs and massive galaxies using the \cite{Landy93} estimator. For the reference, a random catalogs were constructed using background sources ( of redshift range $0.02 > z > 0.1$) detected within the same survey footprint to account for possible incompleteness near bright galaxies. Since both dE subsamples are analyzed using the same reference catalog, any residual systematic effects are expected to affect both measurements similarly and therefore should not significantly influence our differential comparison.

Figure \ref{radprof} shows that round dEs exhibit a significantly stronger clustering signal at small projected separations ($R \lesssim 30\arcmin \simeq 150$ kpc) compared to flat dEs, indicating an excess probability of finding round systems near massive galaxies relative to a random distribution. In contrast, flat dEs show a relatively weak correlation signal and approach a nearly uniform distribution at larger separations. 

We note that the sample was selected from the Virgo dE catalog and is not expected to contain a significant 
population of ultra-compact dwarfs or compact ellipticals. Most galaxies lie at projected distances greater than $\sim50$ kpc (see Figure \ref{radprof}) from the nearest massive galaxy, well outside the regime where tidal stripping is expected to efficiently produce compact stellar systems \citep{Bekki03,Du19,Bian25}.

This result suggests that the intrinsic shapes of dwarf ellipticals are closely linked to their local environment within the Virgo cluster. Round dEs preferentially reside in the deep potential wells and strong tidal fields surrounding massive galaxies, where repeated interactions and gravitational heating can isotropize stellar orbits and gradually transform initially flattened systems into more spheroidal galaxies through tidal grinding. In contrast, flat dEs are more common in the outer regions of the cluster, where weaker environmental processing allows them to retain their disk-like structures and likely reflects a more recently accreted population.

\begin{figure}
\includegraphics[width=8.5cm]{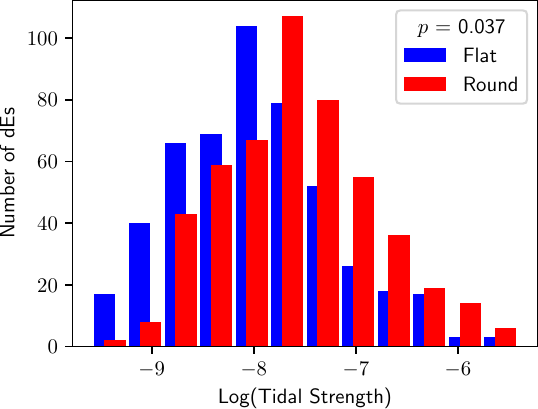}
\caption{Distribution of the projected tidal field strength. Round dEs are preferentially found in stronger tidal environments than flat dEs. The $p$-value from the KS test comparing the two distributions is shown in the legend. }
\label{tfield}
\end{figure}

To characterize the local tidal environment of each dE without assuming an arbitrary cluster center, we compute a projected tidal-field parameter \citep{Byrd90},

\[
T=\sum_i \frac{M_i}{r_i^3},
\]

where \(M_i\) is the stellar mass of each massive galaxy and \(r_i\) is its projected separation from the dE. This quantity serves as a proxy for the cumulative tidal influence exerted by nearby massive galaxies throughout the Virgo cluster. Figure~\ref{tfield} shows the distribution of \(\log T\) for the flat and round dE subsamples. Round dEs are systematically shifted toward higher tidal-field values, indicating that they preferentially inhabit regions where the local tidal environment is stronger.

It is important to note, however, that the tidal-field parameter used here is intended as a relative measure of the present-day local tidal environment rather than a direct estimate of the physical tidal force experienced by a galaxy. In particular, it differs from the integrated tidal force (ITF) employed by \cite{Lokas2010}, which quantifies the cumulative tidal effects accumulated over a galaxy's orbital history. Owing to the lack of orbital information for the observed galaxies, such an estimate is not possible for our sample. Our results therefore indicate that round dEs preferentially reside in stronger present-day tidal environments, but do not directly measure the total tidal processing experienced by each galaxy over their lifetimes.

A KS test confirms that the two distributions are statistically distinct ($p = 0.037$) within 95\% confidence limit. This result supports a scenario in which repeated tidal interactions and gravitational heating progressively transform initially flattened systems into rounder spheroids through tidal grinding, while flat dEs likely represent a more recently accreted population that has experienced weaker environmental processing.

\subsection{Kinematics}
\begin{figure}
\includegraphics[width=8.5cm]{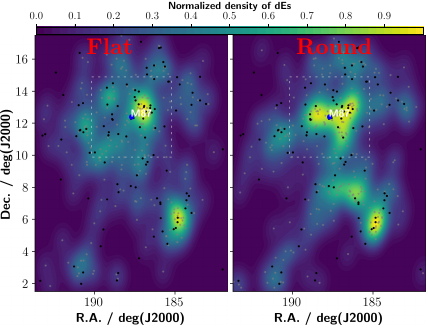}
\caption{Surface number density maps of dEs with available radial velocities in the Virgo cluster. The symbols and color scheme are the same as in Figure \ref{classmap}. The central $5\arcdeg\times\,5\arcdeg$ region is highlighted, corresponding to the area used to examine the line-of-sight velocity distributions of flat and round dEs within the cluster core. }
\label{classmap_z}
\end{figure}

Although our primary analysis uses the full dE sample without requiring radial velocity measurements, a subset of galaxies (498 out of 1108) has spectroscopic velocities available. These measurements were compiled by cross-matching our catalog with publicly available spectroscopic databases, including NED, SDSS, and the recent DESI survey.

This subsample allows us to confirm Virgo cluster membership and assess possible contamination from foreground or background objects. Using this subsample, we construct KDE maps of flat and round dEs (Figure \ref{classmap_z}) and find that the spatial trends remain consistent with those derived from the full sample, despite the reduced statistics. Round dEs are strongly concentrated around the massive galaxies of the Virgo cluster tracing the deepest local potential wells. In contrast, flat dEs exhibit a more spatially extended and homogeneous distribution across the cluster, with only a single prominent concentration near the cluster core around M87. The persistence of this segregation in the spectroscopic subsample suggests that it is not driven by projection effects, but instead reflects different evolutionary histories for flat and round dEs within the cluster environment.

\begin{figure}
\includegraphics[width=8.5cm]{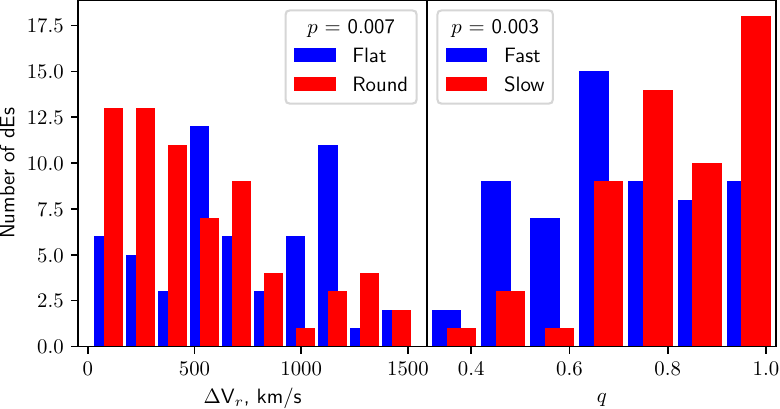}
\caption{Left: Distribution of line-of-sight velocities of dEs relative to M87. The vertical dashed lines indicate the velocity thresholds used to define the slow- and fast-moving subsamples. Right: Projected axis-ratio distributions of slow and fast dEs, defined as galaxies with relative line-of-sight velocities below 475\,$\kms$ and above 555\,$\kms$, respectively. The corresponding KS-test p-value is indicated in the legend. Fast-moving dEs exhibit a larger fraction of flattened systems, whereas slow-moving dEs are preferentially round, consistent with a morphology--kinematics relation within the Virgo cluster.}
\label{velprof}
\end{figure}

Previous work by L09 showed that flat and round dEs in the Virgo cluster differ not only in their spatial distributions but also in their kinematic properties: flat systems tend to move faster relative to the cluster mean velocity, while round systems move more slowly. Their analysis, however, was limited to only 46 dEs with available radial velocities located in the central $1^\circ \times 1^\circ$ region of the Virgo cluster.

The Virgo cluster is dynamically complex and contains several substructures, including the M87, M49, and NGC~4261 groups, each with distinct systemic velocities. To minimize the impact of these substructures, we focus on the cluster core, the central $5\arcdeg\times\,5\arcdeg$ or , region highlighted in Figure~\ref{classmap_z}. Within this region we identify 149 dEs with available radial velocity measurements, selected using a velocity cut of $\pm1500~\kms$ relative to the line-of-sight velocity of M87, i.e., $v_r = 1284~\kms$. This sample provides a statistically meaningful dataset for examining the kinematic properties of dEs with respect to the cluster center.

Figure~\ref{velprof}, left panel, shows the distribution of line-of-sight velocities of dEs relative to M87 ($\Delta v_r$). We find a clear dependence of galaxy shape on velocity: round dEs are more common at lower relative velocities (with median $\Delta v = 414~\kms$), while flat dEs dominate at higher velocities (with median $\Delta v = 654~\kms$). A Kolmogorov--Smirnov (KS) test comparing the velocity distributions of flat and round dEs yields $p = 0.02$, indicating a statistically significant difference between the two populations.

The transition between these regimes occurs near $v_r \approx 550~\kms$. We therefore divide the sample into slow ($v_r < 475~\kms$, 45th percentile) and fast ($v_r > 555~\kms$, 55th percentile) dEs and compare their axis-ratio distributions. As shown in Figure~\ref{velprof} (right panel), fast-moving galaxies preferentially occupy the low-$q$ regime and are therefore more flattened, whereas slow-moving galaxies are predominantly round and populate the high-$q$ region. This kinematic--morphological correlation suggests that round dEs are more dynamically relaxed within the cluster potential, while flatter systems may represent more recently accreted galaxies that have not yet been fully dynamically processed.

\section{Conclusions and Discussions }

Our results provide new evidence that the structural properties of dEs in the Virgo cluster are closely linked to tidal processing within the cluster potential. We find that round dEs are preferentially concentrated near to massive member of Virgo cluster, where tidal forces and repeated high-speed encounters are strongest, while flat dEs exhibit a more extended spatial distribution and larger relative velocities. This pattern is naturally explained by tidal grinding, in which repeated tidal interactions and gravitational heating progressively transform initially flattened, disk-like systems into dynamically hotter, rounder spheroids \citep{Lokas2010,Lokas11,Lokas2020}. In this framework, round dEs represent a dynamically evolved population that is likely to have undergone more substantial cumulative tidal processing over multiple orbits, whereas flat dEs are consistent with a less tidally processed population that has retained a more flattened morphology.

This interpretation is consistent with the findings of \citet{Lisker07}, who showed from a quantitative analysis of 413 Virgo cluster dEs that ordinary nucleated dEs are spheroidal objects centrally clustered like elliptical and S0 galaxies, indicating that they have resided in the cluster for a long time or were formed along with it, while dE subclasses shaped like thick disks show no such central clustering and probably formed from infalling progenitor galaxies. Crucially, L09 demonstrated that fast-moving dEs near the Virgo cluster centre are relatively flat objects while slow-moving dEs are nearly round, and that when subdivided by axial ratio, the flat dEs have a broad line-of-sight velocity distribution whereas round dEs show a narrow single peak, a result which closely mirrors our own findings. Further analysis of stellar populations through spectroscopic observations of these dEs would reinforce this picture, as \citet{Paudel10} showed using VLT/FORS2 spectroscopy that nucleated dEs without discs --- which are distributed in regions of higher local density --- are systematically older than dEs with discs, and \citet{Toloba09} found that rotationally supported dEs have, on average, younger stellar populations than pressure-supported systems. 

In this work, we explicitly excluded compact stellar systems such as ultra-compact dwarfs (UCDs) and compact ellipticals (cEs; e.g., M32-like galaxies) from our sample selection. We acknowledge that a small number of dEs in our sample have effective radii comparable to those of the largest compact systems, although these objects do not exhibit the high surface brightness characteristics typically associated with UCDs or cEs. To assess whether such objects influence our results, we repeated the entire analysis after excluding all galaxies with $R_{\rm e}<300$ pc. The resulting spatial and kinematic distributions remain unchanged within the uncertainties, and all reported statistical significances are qualitatively unaffected.

\subsection{In the Context of Numerical Simulations}
The observed segregation extends beyond a simple morphology--density relation. Round dEs are not only concentrated near massive galaxies but also exhibit lower relative velocities and occupy more virialized regions of projected phase space. Any successful formation scenario must therefore reproduce both the environmental and kinematic differences simultaneously. Tidal-processing models naturally predict such behavior because galaxies that spend longer within the cluster potential experience stronger cumulative tidal interactions and greater orbital relaxation \citep[e.g.,][]{Moore96,Mayer01,Mastropietro05}. Consistent with this picture, \citet{Lokas2020} used IllustrisTNG-100 simulations to show that the degree of tidal evolution is primarily governed by the integrated tidal force experienced along a galaxy's orbit. Galaxies that undergo multiple pericentric passages experience substantial mass loss, become gas poor, and show progressive morphological transformation accompanied by a reduction in rotational support. The spatial and kinematic segregation observed in Virgo is therefore consistent with a tidal evolutionary sequence in which round, slowly moving dEs represent the most strongly processed systems, while flat dEs correspond to a less evolved or more recently accreted population.

We do not, however, attempt a direct quantitative comparison between the observed axis-ratio distribution and specific simulation predictions, as this would require tracking galaxy shape evolution as a function of orbital history, tidal field strength, and progenitor structure within a Virgo-like environment. Our results should therefore be interpreted as evidence for a morphology–environment relation consistent with tidal processing, rather than a direct measurement of its efficiency.

\subsection{Cosmological assembly bias}

An alternative interpretation of our results can be discussed with cosmological assembly bias. In the hierarchical $\Lambda$CDM framework, galaxies that assemble earlier or are accreted into dense environments at earlier times can exhibit systematically different structural and dynamical properties from systems of similar mass that form later \citep{Gao05,Wechsler06,Wechsler18}. If the round dEs represent an earlier-accreted population, while the flat dEs are more recent arrivals to the cluster, the observed trend could arise without requiring substantial tidal transformation after infall. In this picture, round dEs would preferentially trace the oldest cluster population, whereas flatter, higher-angular-momentum systems would be associated with later-forming or later-accreted galaxies \citep[e.g.,][]{Lisker09,Janz14,Janssen16}.

\begin{figure}
\includegraphics[width=8.5cm]{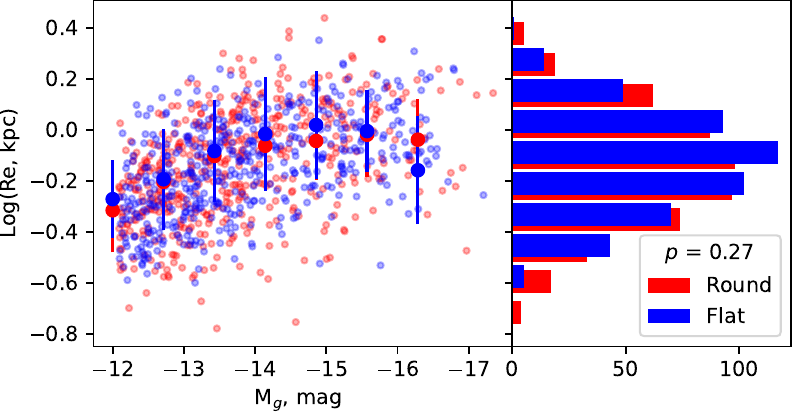}
\caption{Relation between effective radius and absolute magnitude for Virgo dEs. Flat and round dEs are shown in blue and red, respectively. Large symbols indicate median binned values, with error bars showing the standard deviation. Despite their markedly different spatial and kinematic distributions, the two populations exhibit similar size--luminosity relations. A KS test yields $p=0.27$ (left panel), indicating no significant difference in their size distributions. }
\label{sizemag}
\end{figure}

To assess this possibility, we compared the structural properties of the flat and round dE populations. Figure~\ref{sizemag} shows the relation between effective radius and absolute magnitude for the two samples. The half-light radii ($R_e$) were derived from curve-of-growth analysis of $g$-band Legacy Survey imaging, following \citep{Paudel23} Despite their markedly different spatial and kinematic distributions, flat and round dEs follow nearly identical size--luminosity relations. A KS test yields $p=0.27$, indicating no statistically significant difference in their effective-radius distributions. The same result is obtained when only non-nucleated dEs are considered.

While we do not rule out the role of assembly-bias origin, we suggests that the round dEs are not simply a more compact, earlier-formed population. Instead, the observed differences between flat and round dEs appear to be linked primarily to their dynamical and environmental histories rather than to global structural scaling relations.

\subsection{Effect of projection}

The projected axis ratio $q$ does not uniquely determine a galaxy's intrinsic three-dimensional shape. In particular, intrinsically flattened systems viewed close to face-on can appear round in projection, whereas galaxies with very low projected axis ratios are necessarily intrinsically flattened. As a result, the round subsample is expected to contain some contamination from face-on flat galaxies, while the flat subsample remains comparatively pure.

 To quantify the impact of projection effects on our axis-ratio classification, we performed a Monte Carlo simulation following the formalism of \citet{Janssen16}. We modeled the intrinsic three-dimensional shapes of dEs as randomly oriented triaxial ellipsoids characterized by an ellipticity $E = 1 - C/A$ and triaxiality $T = (A^{2}-B^{2})/(A^{2}-C^{2})$, where $A>B>C$ are the intrinsic principal axes. Motivated by the results of \citet{Janssen16}, who found that Virgo dEs are well described by mildly prolate intrinsic shapes, we adopted intrinsic shape distributions centered on $E=0.4\pm0.1$ (corresponding to an intrinsic axis ratio $C/A \simeq 0.6$ which is median value of our flat sample) and $T=0.16\pm0.05$. This choice allows us to estimate the fraction of intrinsically flattened systems that would be classified as round solely due to projection effects.

\begin{figure}
\includegraphics[width=8.5cm]{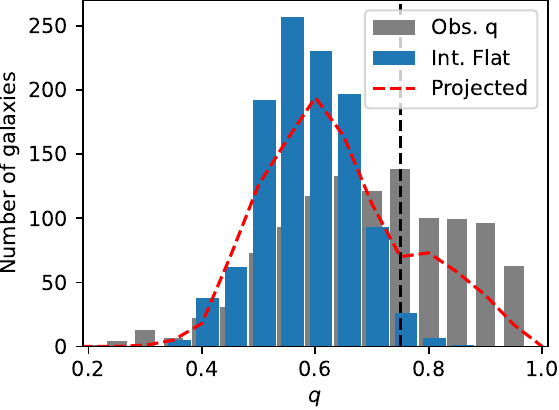}
\caption{Distribution of projected axis ratios for the Virgo dE sample. The observed distribution is shown in gray. The blue histogram represents the assumed intrinsically flat axis-ratio distribution, modeled as a population of triaxial galaxies with median intrinsic axis ratio $C/A = 0.6 \pm 0.1$ and triaxiality $T = 0.16 \pm 0.05$, following the formalism of \citet{Janssen16}. The red curve shows the projected axis-ratio distribution obtained from Monte Carlo realizations of the intrinsically flat population viewed from random orientations.  }
\label{axrsim}
\end{figure}

The resulting projected axis-ratio distribution is shown in Figure~\ref{axrsim} (red curve). We find that $\sim$40\% of intrinsically flattened galaxies with $C/A \simeq 0.6$ would be classified as round ($q>0.74$) due to projection effects. The simulated distribution broadly reproduces the high-$q$ tail of the observed axis-ratio distribution, indicating that a substantial fraction of apparently round dEs may arise from intrinsically flattened systems viewed at favorable orientations. Thus, the observed axis ratio alone does not uniquely determine a galaxy's intrinsic shape. Nevertheless, the flat and round subsamples exhibit statistically significant differences in their spatial and kinematic distributions, implying that the observed morphology--environment relation persists despite substantial projection contamination. We therefore regard the Monte Carlo experiment primarily as a demonstration of the importance of projection effects and the associated uncertainty in interpreting galaxy shapes.

A KS test comparing the observed and simulated axis-ratio distributions yields $p = 7.5\times10^{-28}$, indicating that the projection-only model does not fully reproduce the observed axis-ratio distribution. Thus, although projection effects substantially contaminate the round subsample, they are unlikely to be the sole origin of the observed high-$q$ population.

\newpage
\begin{acknowledgments}
We thank Harry Ferguson for fruitful discussions and comments on the draft version of this paper.
SP and SJY acknowledge support from the Mid-career Researcher Program (RS-2023-00208957 and RS-2024-00344283, respectively) through Korea's National Research Foundation (NRF). 
SJY acknowledge support from the Basic Science Research Program (2022R1A6A1A03053472) through Korea's NRF funded by the Ministry of Education. 

The DESI Legacy Imaging Surveys consist of three individual and complementary projects: the Dark Energy Camera Legacy Survey (DECaLS), the Beijing-Arizona Sky Survey (BASS), and the Mayall z-band Legacy Survey (MzLS). DECaLS, BASS and MzLS together include data obtained, respectively, at the Blanco telescope, Cerro Tololo Inter-American Observatory, NSF’s NOIRLab; the Bok telescope, Steward Observatory, University of Arizona; and the Mayall telescope, Kitt Peak National Observatory, NOIRLab. NOIRLab is operated by the Association of Universities for Research in Astronomy (AURA) under a cooperative agreement with the National Science Foundation. Pipeline processing and analyses of the data were supported by NOIRLab and the Lawrence Berkeley National Laboratory (LBNL). Legacy Surveys also uses data products from the Near-Earth Object Wide-field Infrared Survey Explorer (NEOWISE), a project of the Jet Propulsion Laboratory/California Institute of Technology, funded by the National Aeronautics and Space Administration. Legacy Surveys was supported by: the Director, Office of Science, Office of High Energy Physics of the U.S. Department of Energy; the National Energy Research Scientific Computing Center, a DOE Office of Science User Facility; the U.S. National Science Foundation, Division of Astronomical Sciences; the National Astronomical Observatories of China, the Chinese Academy of Sciences and the Chinese National Natural Science Foundation. LBNL is managed by the Regents of the University of California under contract to the U.S. Department of Energy. The complete acknowledgments can be found at https://www.legacysurvey.org/acknowledgment/.

This research used data obtained with the Dark Energy Spectroscopic Instrument (DESI). DESI construction and operations is managed by the Lawrence Berkeley National Laboratory. This material is based upon work supported by the U.S. Department of Energy, Office of Science, Office of High-Energy Physics, under Contract No. DE–AC02–05CH11231, and by the National Energy Research Scientific Computing Center, a DOE Office of Science User Facility under the same contract. Additional support for DESI was provided by the U.S. National Science Foundation (NSF), Division of Astronomical Sciences under Contract No. AST-0950945 to the NSF’s National Optical-Infrared Astronomy Research Laboratory; the Science and Technology Facilities Council of the United Kingdom; the Gordon and Betty Moore Foundation; the Heising-Simons Foundation; the French Alternative Energies and Atomic Energy Commission (CEA); the National Council of Science and Technology of Mexico (CONACYT); the Ministry of Science and Innovation of Spain (MICINN), and by the DESI Member Institutions: www.desi.lbl.gov/collaborating-institutions. The DESI collaboration is honored to be permitted to conduct scientific research on Iolkam Du’ag (Kitt Peak), a mountain with particular significance to the Tohono O’odham Nation. Any opinions, findings, and conclusions or recommendations expressed in this material are those of the author(s) and do not necessarily reflect the views of the U.S. National Science Foundation, the U.S. Department of Energy, or any of the listed funding agencies.

Funding for SDSS-III has been provided by the Alfred P. Sloan Foundation, the Participating Institutions, the National Science Foundation, and the U.S. Department of Energy Office of Science. The SDSS-III web site is http://www.sdss3.org/.

SDSS-III is managed by the Astrophysical Research Consortium for the Participating Institutions of the SDSS-III Collaboration including the University of Arizona, the Brazilian Participation Group, Brookhaven National Laboratory, Carnegie Mellon University, University of Florida, the French Participation Group, the German Participation Group, Harvard University, the Instituto de Astrofisica de Canarias, the Michigan State/Notre Dame/JINA Participation Group, Johns Hopkins University, Lawrence Berkeley National Laboratory, Max Planck Institute for Astrophysics, Max Planck Institute for Extraterrestrial Physics, New Mexico State University, New York University, Ohio State University, Pennsylvania State University, University of Portsmouth, Princeton University, the Spanish Participation Group, University of Tokyo, University of Utah, Vanderbilt University, University of Virginia, University of Washington, and Yale University.

\end{acknowledgments}

\facilities{DESI, SDSS, CDS, NED}
\software{The following software tools were utilized in this work: Astropy \cite{astropy13, astropy18, astropy22}, Matplotlib \cite{matplotlib}, Numpy \cite{numpy}, SciPy \cite{2020SciPy}}

\newpage
%\bibliography{myref}{}
%\bibliographystyle{aasjournal}

\end{document}